\documentclass[pre,floatfix,twocolumn,showpacs,a4]{revtex4}

\usepackage{graphicx}
\usepackage{psfrag} 
\usepackage{epsfig}    
\usepackage{rotating}  

\begin{document}

\title{Dynamics of latent voters}
\author{Renaud Lambiotte$^{1,2}$\email{Renaud.Lambiotte@uclouvain.be}}
\author{Jari Saram\"aki$^{3}$}
\author{Vincent D. Blondel$^{1}$}
\affiliation{
  $^1$ Dept.~of Mathematical Engineering, Universit\'e catholique de Louvain,  4 avenue Georges Lemaitre,
B-1348 Louvain-la-Neuve, Belgium \\
$^2$ Institute for Mathematical Sciences, Imperial College London, 53 Prince's Gate, South Kensington campus, SW7 2PG, UK\\
$^{2}$Dept.~of Biomedical Engineering and Computational Science, Helsinki University of Technology, P.O. Box 9203, FI-02015 TKK, Finland 
}

\date{\today}

\begin{abstract}
We study the effect of latency on binary-choice opinion formation models. Latency is introduced into the models as an additional dynamic rule: after a voter changes its opinion, it enters a waiting period of stochastic length where no further changes take place. We first focus on the voter model and show that as a result of introducing latency, the average magnetization is not conserved, and the system is driven toward zero magnetization, independently of initial conditions. The model is studied analytically in the mean-field case and by simulations in one dimension. We also address the behavior of the Majority Rule model with added latency, and show that the competition between imitation and latency leads to a rich phenomenology.

\end{abstract}
\pacs{89.75.-k, 02.50.Le, 05.50.+q, 75.10.Hk}

\maketitle

\section{Introduction}

Binary-choice opinion formation models address the emergence of ordered (disordered) states, i.e.~consensus or coexistence of different opinions~\cite{review,bfootnote}. In these models, opinions of agents are influenced by those of other agents. This influence is captured by dynamic rules which are iterated until some stable state is reached. 
The \emph{voter model}~\cite{orig,review} can be considered as an archetypal binary opinion formation model. It has found applications in many fields going from social dynamics and population genetics to chemical kinetics. The model consists of $N$ agents, each endowed with a binary variable $s = \uparrow$ or $s = \downarrow$. The agents can be fully mixed, so that each agent can interact with all the others, or located on a lattice or network, which then mediates the interactions. At each time step, an agent $i$ is selected along with one of its neighbors $j$ and the agent adopts the opinion of the neighbor, $s_i=s_j$. By construction, agents therefore imitate their neighbors and are subject to peer pressure in the average sense, changing their opinion with a probability equal to the fraction of neighbors that disagree with them. This simple  rule implies that the average opinion, analogous to magnetization in spin models,  is conserved~\cite{footnote_networks}. This renders the voter model soluble in all dimensions \cite{L99,K02} and makes it a paradigmatic model for the emergence of an ordered state in a non-equilibrium system. 

Over the last years, the voter model has gained much attention in the physics community, especially regarding the role of different interaction network topologies \cite{sood,sw}, e.g. scale-free, small-world, etc., on the opinion dynamics.
Several generalizations have also been proposed in order to allow for more complicated agent interactions, such as the Majority Rule \cite{Galam,krap}, non-linear voter models \cite{nonlinear, nonlinear2,vacillating} or multi-state voter models~\cite{Sire1995,Xavi}. In most of these models, agents interact with each other whatever their history, namely, at each time step of the dynamics, voters change opinion with a probability that depends only on the configuration of their neighbors. Hence, the opinions of individual agents can be very volatile over time. Note that there are also models which consider the spreading of e.g.~a fashion over a susceptible agent population, such as the threshold model~\cite{Watts2002}, where the state of an agent is frozen once it has adopted the opinion. Here, we study opinion formation dynamics between these two extremes, by introducing {\em latency} to binary-choice models. Instead of the Markovian assumption where an agent's past choices have no influence to its present, we regulate the frequency of its opinion changes by applying a latency period. Consider e.g.~the adoption of a new technology, such as choosing between a Bluray or a HD-DVD player. It is likely that the choice of a customer is influenced by his acquaintances; however, it is unlikely that the customer will replace his equipment immediately after purchasing - rather, a new acquisition is made only after the previous device is broken or obsolete. Similar memory effects take place in the competition between mobile phone operators, as customers are usually bound to one or two year contracts. In general, memory effects are important in situations where there is some cost or restriction associated with switching opinions.  A better understanding of the effects of memory on opinion formation dynamics is therefore of interest~\cite{castellano,Schweitzer}.

In this paper, we study the effects of a simple modification applicable to several opinion formation models: agents cannot be influenced by their neighbours for some (stochastic) period of time after they change opinion. Note that this bears close resemblance to the immunity time in SIRS models of epidemic spreading (see, e.g.~\cite{Hethcote}).
In Section~\ref{latent}, we incorporate this mechanism to the voter model. The model is first analytically studied in the mean-field case and it is shown that average magnetization is not conserved by the dynamics, contrary to the original voter model, and that latency drives the system toward zero magnetization. In the one dimensional case, computer simulations show that the dynamics also has a tendency toward zero magnetization but that the exit probability, i.e. the probability to reach a consensus state, exhibits a non-trivial  dependence on the initial condition, even in the limit of infinitely large systems. In Section~\ref{mr}, we focus on a generalization of the Majority Rule and show that the competition between imitation and latency leads to much richer phenomenologa. In Section~\ref{conclusion}, finally, we conclude and discuss possible generalizations that might be of interest.

\section{Latent voters}
\label{latent}
We propose a simple variation of the voter model that incorporates a memory for the agent. Here we assume that voters are not only characterized by their opinion $s$, but also by their activity, $I$ (Inactive) or $A$ (Active). The system is governed by the following discrete-time-step rules: 

(i) A random voter is picked (the focal voter); 

(ii) If the focal voter is active ($A$), it adopts the state of a randomly chosen
neighbor with a probability $p$. If this leads to a change of his opinion, the focal voter becomes inactive;

(iii) If the focal voter is inactive ($I$), it is reactivated with probability $q$; 

These steps are repeated {\it ad infinitum} or until consensus is reached.  The model therefore incorporates a latent time  
between two opinion switches of the same agent.
For the sake of simplicity, let us first focus on a fully mixed system (i.e.~a fully connected network), where everybody is connected to everybody and hence a mean-field description is justified. 
Moving to continuous time, it is straightforward to derive the system of equations governing the dynamics. As an example, let us explicitly describe the dynamics of the number $N_{a\uparrow}$ of active nodes of opinion
$\uparrow$. The discrete-time-step equation for $N_{a\uparrow}$ of such nodes can be written as 
\begin{equation}
N_{a\uparrow}\left(t+\Delta t\right)
= N_{a\uparrow}\left(t\right)-p\left(\frac{N-N_\uparrow}{N}\right)N_{a\uparrow}+q\left(N_\uparrow-N_{a\uparrow}\right),
\end{equation}
where $N$ is the system size, and $N_\uparrow$ is the number of nodes of opinion $\uparrow$ irrespective of their status (active/inactive). Taking the continuous time limit, we get
\begin{eqnarray}
\partial_t{N_{a\uparrow}} & = & \lim_{\Delta t\rightarrow 0}\frac{N_{a\uparrow}\left(t+\Delta t\right)-N_{a\uparrow}\left(t\right)}{\Delta t} \cr & = & \lim_{\Delta t\rightarrow 0} -\frac{p}{\Delta t}\left(\frac{N-N_\uparrow}{N}\right)N_{a\uparrow}+\frac{q}{\Delta t}\left(N_\uparrow-N_{a\uparrow}\right). \cr
 & &
\end{eqnarray}
We can now take the limit such that $ \lim_{\Delta t\rightarrow 0} p/\Delta t = 1$, \emph{i.e.}, $p$ determines the time scale. Thus $\lim_{\Delta t\rightarrow 0} q/\Delta t = \lim_{\Delta t\rightarrow 0} q/p \equiv \lambda$ determines the average latent time, $\left<\tau\right>\approx 1/\lambda$. It is now straightforward to show that the system dynamics is determined by the following system of equations:
\begin{eqnarray}
\label{eq1}
\partial_t \rho_{\uparrow} &=& \rho_{a\downarrow} \rho_\uparrow -   \rho_{a\uparrow} (1-\rho_\uparrow) \cr
\partial_t \rho_{a\uparrow} &=& -   \rho_{a\uparrow} (1-\rho_\uparrow) + \lambda (\rho_\uparrow - \rho_{a\uparrow}) \cr
\partial_t \rho_{a\downarrow} &=& -    \rho_{a\downarrow} \rho_\uparrow + \lambda (1-\rho_\uparrow - \rho_{a\downarrow}),
\end{eqnarray}
where $\rho_\uparrow$ is the density of nodes of opinion $\uparrow$ and $\rho_{a\uparrow}$ is the density of active nodes of opinion $\uparrow$, and similarly for ($\downarrow$). In the first equation of (\ref{eq1}), the gain term accounts for situations where the focal node is active with a  $\downarrow$ opinion, while the randomly selected neighbor (whatever his activation) has a $\uparrow$  opinion, and similarly for the loss term. In the second equation of (\ref{eq1}), the loss term accounts for situations where an active $\uparrow$ voter switches opinion, while the gain term comes from the reactivation of voters with rate $\lambda$.

In the limit $\lambda \rightarrow \infty$, voters reactivate infinitely fast and we recover the classical mean field equations for the voter model 
\begin{eqnarray}
\rho_{a\uparrow} &=& \rho_\uparrow,  \cr
 \rho_{a\downarrow} &=&  1-\rho_\uparrow ,
 \end{eqnarray}
and
\begin{eqnarray}
\partial_t \rho_{\uparrow} = (1-\rho_\uparrow) \rho_\uparrow -   \rho_{\uparrow} (1-\rho_\uparrow) =0,
\end{eqnarray}
thereby confirming that the density of $\uparrow$ voters is conserved in the voter model.

When $\lambda$ is finite, in contrast, the density of $\uparrow$ voters is not conserved but it goes to the zero-magnetization solution $\rho_{\uparrow}=1/2$ for any initial condition, with $\rho_{a\uparrow}=\rho_{a\downarrow}=\frac{\lambda}{ (2+\lambda)}$. This  attractor of the dynamics can be found by using the second and third equations of (\ref{eq1}) in order to express $\rho_{a\uparrow}$ and $\rho_{a\downarrow}$ as a function of $\rho_{\uparrow}$
\begin{eqnarray}
  \rho_{a\uparrow} &=& \frac{ \lambda \rho_\uparrow }{(1-\rho_\uparrow + \lambda)} \cr
  \rho_{a\downarrow} &=& \frac{ \lambda (1-\rho_\uparrow) }{(\rho_\uparrow + \lambda)},
\end{eqnarray}
The first equation of (\ref{eq1}) therefore leads to the condition
\begin{eqnarray}
\label{cond}
 \frac{\lambda \rho_{\uparrow} (\rho_{\uparrow} -1) (2 \rho_{\uparrow} -1)}{(1-\rho_\uparrow + \lambda) (\rho_\uparrow + \lambda)} = 0,
\end{eqnarray}
from which one sees that $\rho_{\uparrow} =1/2$, $\rho_{\uparrow} =0$ and $\rho_{\uparrow} =1$ are the stationary solutions of the problem.  It is straightforward to show that the consensus solutions $\rho_{\uparrow}=0$ and $\rho_{\uparrow}=1$ are unstable.
 To do so, let us look at small deviations around the stationary solution $\rho_{\uparrow}=0$: $\rho_{\uparrow}=\epsilon_{\uparrow}+o(\epsilon^2)$, $\rho_{a\uparrow}=\epsilon_{a\uparrow}+o(\epsilon^2)$ and $\rho_{a\downarrow}=1-\epsilon_{a\downarrow}+o(\epsilon^2)$. In that case, the first two equations of Eq.(\ref{eq1}) become
\begin{eqnarray}
\label{eq1Lin}
\partial_t \epsilon_{\uparrow} =  \epsilon_\uparrow -   \epsilon_{a\uparrow}  \cr
\partial_t \epsilon_{a\uparrow} =  \lambda \epsilon_\uparrow - (\lambda +1) \epsilon_{a\uparrow}. 
\end{eqnarray}
This set of linearized equations has one positive eigenvalue for any values of $\lambda$, except in the voter model limit $\lambda \rightarrow \infty$ where both eigenvalues go to zero. One can similarly show that the consensus $\rho_{\uparrow}=1$ is unstable, while $\rho_{\uparrow}=1/2$ is a stable solution for any value of $\lambda$.
 Thus a population is driven away from consensus toward the only stable solution, i.e. the zero-magnetization state.
It should be stressed that such a drift toward zero magnetization also takes place in other models for opinion formation, but due to different mechanisms. For instance, in the vacillating voter model \cite{vacillating}, formation of consensus is hindered by the uncertainty of the agents. In the present model, in contrast, it is only memory that forestalls consensus.  

\begin{figure}[]
\includegraphics[angle=-90,width=0.45\textwidth]{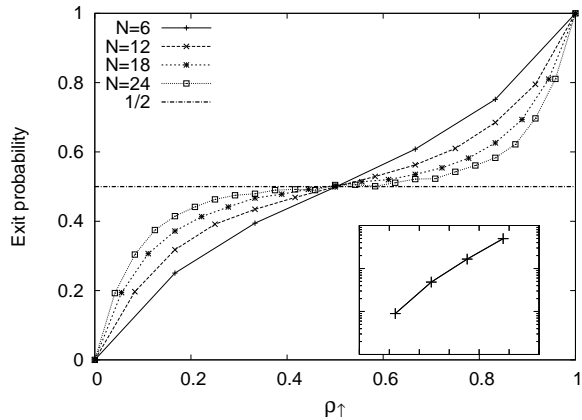}
\vskip .3in
\caption{Exit probability $\mathcal{E}(\rho_{\uparrow})$ versus the density of
$\uparrow$ voters $\rho_{\uparrow}$ for $N=6$, $N=12$, $N=18$ and $N=24$ in the mean-field case. In the inset, we plot the average time to reach consensus as a function of $N$ on a log scale. Simulation results are averaged over 5000 realizations of the random process.}
\label{exit}      
\end{figure}

As for the voter model, the dynamics for a finite population differ from this mean-field description because consensus is ultimately always reached, as it is the only absorbing state of the stochastic dynamics. This implies that the average magnetization does not fluctuate forever around the asymptotic value $\rho_{\uparrow}=1/2$ but that it  asymptotically reaches the state $\rho_{\uparrow}=0$ or $\rho_{\uparrow}=1$, even if those solutions are unstable.  To characterize the evolution to this state, we focus therefore on the exit probability 
$\mathcal{E}(n,N)$, defined as the probability that a population of $N$ voters
ultimately reaches $\uparrow$ consensus when there are initially $n=\rho_{\uparrow} N$
$\uparrow$ voters. Since the density of $\uparrow$ voters and the exit probability are related through 
$\rho_{\uparrow}(\infty)=\mathcal{E}(n)$ and $\rho_{\uparrow}=1/2$ is the only stable solution, it is straightforward to show that the exit probability is equal to
$\mathcal{E}(n,N) =  \frac{1}{2}$ for sufficiently large systems.
The exit probability is therefore independent
of the initial density of $\uparrow$ voters. This is expected because almost all initial states are driven to the potential well at $\rho_{\uparrow}=1/2$ and initial conditions are rapidly ``forgotten" by the dynamics. This is indeed what we observe by performing computer simulations of the model (see Fig.~1), i.e. $\mathcal{E}(\rho_{\uparrow})$ gets closer and closer to $1/2$ when $N$ increases. This interpretation is also confirmed by looking at 
 the time to reach consensus $t_n$.  Computer simulations show that the consensus time scales
exponentially with the number of agents $N$. This anomalously long time is due to the fact that the system has to escape a potential well in order to reach consensus.

\begin{figure}[]
\includegraphics[angle=-90,width=0.45\textwidth]{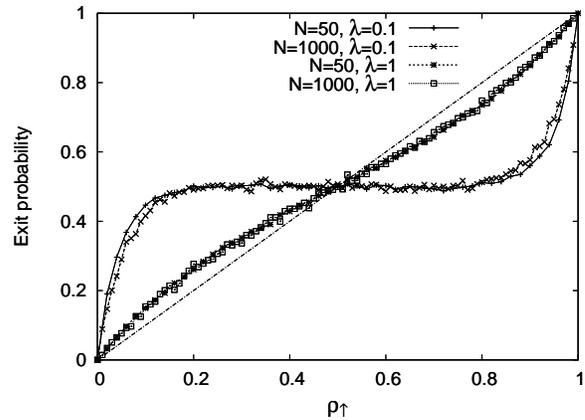}
\vskip .3in
\caption{Exit probability $\mathcal{E}(\rho_{\uparrow})$ versus the density of
$\uparrow$ voters $\rho_{\uparrow}$ for $N=50$ and $N=1000$, in the case of a one-dimensional system. The exit probability has a non-trivial dependence on the initial condition, with a trend toward zero magnetization. When $\lambda$ is decreased, the dynamics loses its dependence on the initial conditions and $\mathcal{E}(\rho_{\uparrow})=1/2$.}
\label{1D}      
\end{figure}

Let us now focus on the latent voter model dynamics on a one-dimensional lattice. In that case, numerical simulations (see Fig. \ref{1D}) show that unlike above, the exit probability does not depend on the number of agents but exhibits 
a non-trivial dependence on the initial conditions. One should stress that such a qualitative change between the mean-field description and one-dimensional dynamics also takes place for other models that do not conserve the average magnetization, such as the majority rule \cite{MR}, the Sznajd model \cite{nonlinear2,sznRel,sznRel2} and non-conservative voter models \cite{vacillating,nonlinear2}.  For small systems ($N=50$), we directly measure
the probability $\mathcal{E}(n,N)$ that the population ultimately reaches a
$\uparrow$ consensus when there are initially $n$ $\uparrow$ voters, averaging over 5000 realizations of the dynamics.  In the case of larger systems ($N=1000$ nodes), we use a different approach by running the dynamics up to $1000$ time steps per agent and
measuring the magnetization at this time.  We then average over 100
realizations of the process to obtain $\rho_{\uparrow}(\infty)$ and finally obtain
$\mathcal{E}(x)$ from $\mathcal{E}(x)=\rho_{\uparrow}(\infty)$.  
 
\section{Majority rule}
\label{mr}
 
In the previous section, we have focused on a generalization of the voter model, a model whose dynamics is particularly trivial in the mean-field case. There are, however, many other models for opinion formation, most of which do not preserve average magnetization. In this section, we will incorporate latency to the majority rule~\cite{Galam,krap,MR} as follows. At each step, the system obeys the following discrete-time-step rules: 

(i) A random voter is picked (the focal voter); 

(ii) If the focal voter is active, two of its neighbors are randomly picked. The focal voter adopts the state of those neighbors with a probability $p$ if they both have identical opinions.  If the focal voter switches opinion, it becomes inactive;

(iii) If the focal voter is inactive, it is reactivated with probability $q$; 

These steps are repeated {\it ad infinitum} or until  consensus is reached.  Similarly to the voter model, for a fully mixed system, moving to continuous time and by introducing $\lambda=q/p$, it is straightforward to show that the system dynamics is determined by:
\begin{eqnarray}
\label{eqMR}
\partial_t \rho_{\uparrow} &=& \rho_{a\downarrow} \rho_\uparrow^2 -   \rho_{a\uparrow} (1-\rho_\uparrow)^2 \cr
\partial_t \rho_{a\uparrow} &=& -   \rho_{a\uparrow} (1-\rho_\uparrow)^2 + \lambda (\rho_\uparrow - \rho_{a\uparrow}) \cr
\partial_t \rho_{a\downarrow} &=& -    \rho_{a\downarrow} \rho_\uparrow^2 + \lambda (1-\rho_\uparrow - \rho_{a\downarrow}), 
\end{eqnarray}
where unnecessary constants have been absorbed in the time scale.
In the limit $\lambda \rightarrow \infty$ of infinitely fast reactivation, one finds a closed equation for $\rho_{\uparrow}$:
\begin{eqnarray}
\partial_t \rho_{\uparrow} &=& (1-\rho_\uparrow) \rho_\uparrow^2 -   \rho_{\uparrow} (1-\rho_\uparrow)^2 \cr&=& - (\rho_\uparrow-1) \rho_\uparrow (2 \rho_\uparrow -1),
\end{eqnarray}
whose only stable solutions are easily seen to be consensus $\rho_{\uparrow=0}$ or $\rho_{\uparrow}=1$.

\begin{figure}[]
\includegraphics[angle=-90,width=0.45\textwidth]{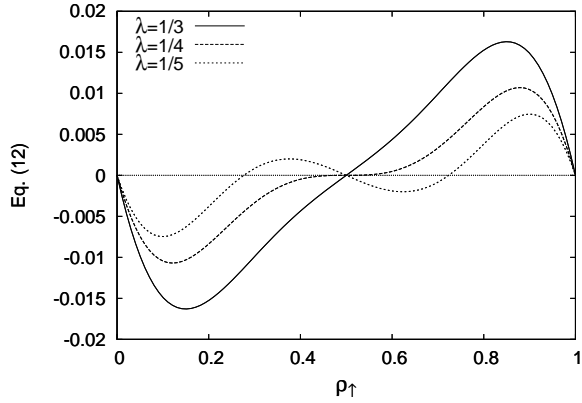}
\vskip .3in
\caption{Eq.~(12) as a function fo the density of $\uparrow$ voters $\rho_{\uparrow}$ when $\lambda=1/3$, $\lambda=1/4$ and
$\lambda=1/5$. When $\lambda>1/4$, the curve has three zeros, two of which correspond to stable solutions $\rho_{\uparrow}=0$ and $\rho_{\uparrow}=1$. When $\lambda<1/4$, the curve has five zeros, three of them corresponding to stable solutions $\rho_{\uparrow}=0$, $\rho_{\uparrow}=1$, and $\rho_{\uparrow}=1/2$.  }
\label{fig2}      
\end{figure}

When $\lambda$ is finite, in contrast, the system exhibits a competition between two opposite effects, i.e. ~an attraction toward consensus due to the majority rule and a tendency toward zero magnetization due to memory effects. Thus one may expect that the system exhibits critical phenomena for some value of $\lambda$. The stationary solutions of (\ref{eqMR}) are easily found by expressing $\rho_{a\uparrow}$ and $\rho_{a\downarrow}$ as a function of $\rho_{\uparrow}$:
\begin{eqnarray}
  \rho_{a\uparrow} &=& \frac{ \lambda \rho_\uparrow }{((1-\rho_\uparrow)^2 + \lambda)} \cr
  \rho_{a\downarrow} &=& \frac{ \lambda (1-\rho_\uparrow) }{(\rho_\uparrow^2 + \lambda)}.
\end{eqnarray}
The first equation of (\ref{eqMR}) leads to the condition
\begin{eqnarray}
\label{cond}
 \frac{\lambda \rho_{\uparrow} (1-\rho_{\uparrow} ) (2 \rho_{\uparrow} -1) (\lambda - \rho_{\uparrow} + \rho_{\uparrow}^2)}{(\rho_\uparrow^2 + \lambda) (1 + \lambda - 2 \rho_\uparrow + \rho_\uparrow^2)} = 0,
\end{eqnarray}
whose stationary solutions may be zero magnetization $\rho_{\uparrow} =1/2$, consensus $\rho_{\uparrow} =0$ and $\rho_{\uparrow} =1$, or $\rho_{\uparrow} = (1 \pm \sqrt{1-4 \lambda})/2$ (see Fig.~2). The system clearly exhibits a qualitative change at $\lambda=1/4$. To show so, let us perform stability analysis of the zero magnetization solution $\rho_{\uparrow} =1/2$. The largest eigenvalue is

\begin{eqnarray}
r= \frac{- (1-4 \lambda)^2 + \sqrt{(1-4 \lambda) (-1 + 44 \lambda + 80 \lambda^2 + 64 \lambda ^3)}}{8 (1+ 4 \lambda)}.\cr
\end{eqnarray}
One finds that the real part of $r$ is positive when $\lambda>1/4$, which implies that the zero magnetization solution is stable only when $\lambda<1/4$. This is expected, as one knows that $\rho_{\uparrow} =1/2$ is unstable when $\lambda \rightarrow \infty$. The stability analysis of the consensus solution $\rho_{\uparrow} =0$ and $\rho_{\uparrow} =1$ leads to a more surprising  result, as it shows that consensus is stable for any value of $\lambda$. One can also show that $\rho_{\uparrow} = (1 \pm \sqrt{1-4 \lambda})/2$ is always unstable. Thus the system exhibits a transition from a regime where  only consensus is a stable solution ($\lambda>1/4$) to a regime where consensus and zero magnetization are stable solutions ($\lambda<1/4$). In that case, the system may reach either of those solutions depending on its initial condition. 

\begin{figure}[]
\includegraphics[angle=-90,width=0.45\textwidth]{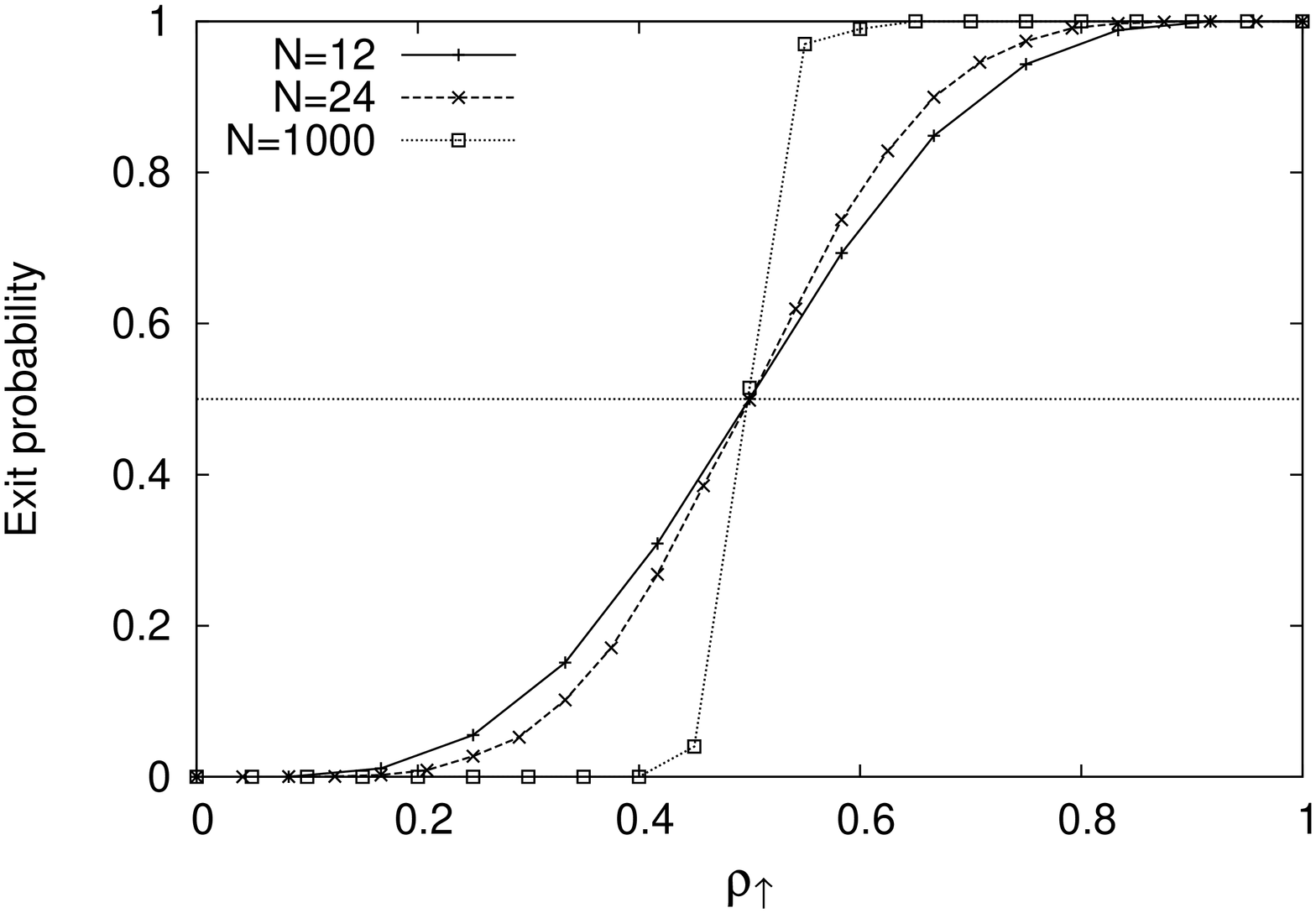}
\includegraphics[angle=-90,width=0.45\textwidth]{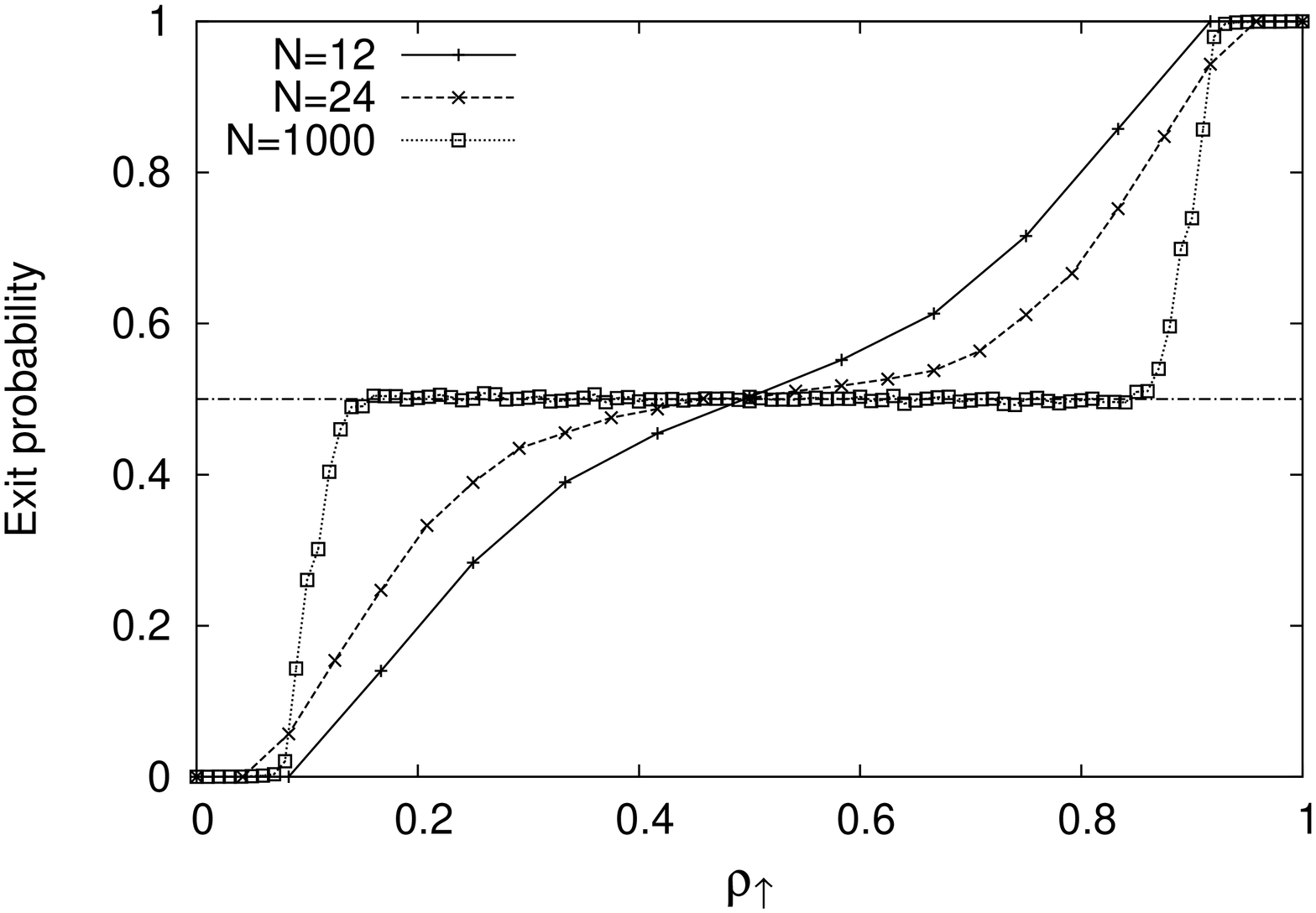}
\vskip .3in
\caption{Exit probability $\mathcal{E}(\rho_{\uparrow})$ versus the density of
$\uparrow$ voters $\rho_{\uparrow}$ for the majority rule with latency. The system is composed of $N=12$, $N=24$ and $N=1000$ agents. In the upper panel, $\lambda=1/2$ and only consensus is a stationary solution. In the lower panel, in contrast, $\lambda=1/20$ and the system may reach consensus or zero-magnetization depending on its initial condition. When $N$ is increased, the exit probability approaches a stepwise function.}
\label{exit}      
\end{figure}

Before concluding, one should stress that the transition that we observe differs from the usual order-disorder transition that would take place if noise was added to the system. Indeed, let us consider a system without memory where agents may perform majority steps with probability $(1-q)$ or switch their state randomly with probability $q$. In this case, the system also exhibits a competition between a mechanism that drives the system to zero magnetization, i.e. the random flips, and a mechanism that drives it towards consensus, i.e. Majority Rule. It is easy to show that this system also exhibits a transition at $q_c=1/3$, but the transition is now from a disordered state to an ordered state. Above $q_c$, the only stable solution is zero magnetization. Below $q_c$, in contrast, zero magnetization ceases to be stable and the system goes to the ordered solution $\rho_{\uparrow}=1/2 \pm \sqrt{(1-3 q)/(1-q)}$. 

\section{Conclusions}
\label{conclusion}

To summarize, we have studied the effects of latency on binary-choice opinion formation models. The motivation for introducing
latency has been to strive for more realistic "sociodynamic" models of situations where there are costs or restrictions associated with
switching opinions, limiting the frequency of opinion changes. We have  shown that a simple 
additional dynamic rule representing latency leads to rich, non-trivial behavior. In the Voter model, 
magnetization is not conserved -- instead, the system is driven towards zero magnetization. Computer simulations indicate an anomalously long consensus time for finite-size systems as the system has to escape a potential well; although
consensus is ultimately reached, the two opinions coexist in the system for very long times ($\sim \exp(N)$). 
For finite-size one-dimensional systems, the exit probability shows a non-trivial dependence
on the initial condition. For the Majority Rule model, introducing latency leads to a rich behavior: depending on the
latency period, the system exhibits a transition from a consensus regime to a regime where both consensus and
zero magnetization are stable solutions. 

We expect that the proposed latency rule is likely to significantly alter the dynamics of other opinion
formation models as well~\cite{nonlinear2,sznRel,sznRel2,vacillating}.
Furthermore, 
a natural next step is to study the effects of latency beyond the mean-field or 1-D-cases:
is even richer behavior to be found for higher-dimensional systems, or when complex networks are mediating the interactions? 
For complex networks, it would be also interesting to study the effects that more realistic network structures have on the latent voting dynamics, e.g.~networks with
community structure~\cite{Castello2007,com}.
We hope that the above results for the simplest opinion formation models and interaction topologies will stimulate further research.

\acknowledgments We gratefully acknowledge the support of the ARC ``Large
Graphs and Networks'' (VB and RL) and the Academy of Finland, the Finnish Center of Excellence Program, Project No.~213470 (JS). This work has been initiated during a visit of JS at Louvain-la-Neuve supported by the ARC ``Large
Graphs and Networks''.

\end{document}